\documentclass[twocolumn,secnumarabic,amssymb,nobibnotes,aps,superscriptaddress,prl]{revtex4-2}

\usepackage{graphicx}
\usepackage{dcolumn}
\usepackage{bm}

\usepackage{xcolor}
\usepackage{amssymb}
\usepackage{wasysym}

\newcommand{\thCsCo}{$\theta$-CsCo}
\newcommand{\thCsZn}{$\theta$-CsZn}
\newcommand{\thRbZn}{$\theta$-RbZn}
\newcommand{\thTlZn}{$\theta$-TlZn}

\newcommand{\jmb}[1]{\textcolor{blue}{#1}}

\begin{document}

\title{Involvement of structural dynamics in the charge-glass formation in molecular metals}

\author{Tatjana Thomas}
\affiliation{Institute of Physics, Goethe University Frankfurt, 60438 Frankfurt (M), Germany}
\author{Yohei Saito}
\affiliation{Institute of Physics, Goethe University Frankfurt, 60438 Frankfurt (M), Germany}
\author{Yassine Agarmani}
\affiliation{Institute of Physics, Goethe University Frankfurt, 60438 Frankfurt (M), Germany}
\author{Tim Thyzel}
\affiliation{Institute of Physics, Goethe University Frankfurt, 60438 Frankfurt (M), Germany}
\author{Kenichiro Hashimoto}
\affiliation{Department of Advanced Materials Science, University of Tokyo, 277-8561 Chiba, Japan}
\affiliation{Institute for Materials Research, Tohoku University, 980-8577 Sendai, Japan}
\author{Takahiko Sasaki}
\affiliation{Institute for Materials Research, Tohoku University, 980-8577 Sendai, Japan}
\author{Michael Lang}
\affiliation{Institute of Physics, Goethe University Frankfurt, 60438 Frankfurt (M), Germany}
\author{Jens M\"uller}
\email[Email: ]{j.mueller@physik.uni-frankfurt.de}
\affiliation{Institute of Physics, Goethe University Frankfurt, 60438 Frankfurt (M), Germany}

\date{\today}

\begin{abstract}
We present a combined study of thermal expansion and resistance fluctuation spectroscopy measurements exploring the static and dynamic aspects of the charge-glass formation in the quasi-two-dimensional organic conductors $\theta$-(BEDT-TTF)$_2$$MM^\prime$(SCN)$_4$ with $M$ = Cs and $M^\prime$ = Co,Zn. In these materials, the emergence of a novel charge-glass state so far has been interpreted in purely electronic terms by considering the strong frustration of the Coulomb interactions on a triangular lattice. 
Contrary to this view, we provide comprehensive evidence for the involvement of a \textit{structural} glass-like transition at $T_{\text{g}} \sim 90-100\,$K.  This glassy transition can be assigned to the freezing of structural conformations of the ethylene endgroups in the donor molecule with an activation energy of $E_{\rm{a}}\approx 0.32\,$eV, and the concomitant slowing down of the charge carrier dynamics is well described by a model of non-exponential kinetics.
These findings discolse an important aspect of the phase diagram and renders the current understanding of the charge-glass state in the whole family of $\theta$-(BEDT-TTF)$_2MM^\prime$(SCN)$_4$ incomplete. Our results suggest that the entanglement of slow structural and charge-cluster dynamics due to the intimate coupling of lattice and electronic degrees of freedom determine the charge-glass formation under geometric frustration.

\end{abstract}

\maketitle

When a first-order phase transition is kinetically avoided by rapid cooling, a new state with different physical properties can emerge \cite{Kagawa2017}. Such a non-equilibrium state, which can be induced by quenching the system that -- in thermal equilibrium -- exhibits a charge-ordering (CO) transition due to strong electronic correlations, was recently discovered in the quasi-two-dimensional organic conductors $\theta$-(BEDT-TTF)$_2$$X$ \cite{Kagawa2013,Sato2014a,Sato2014,Sasaki2017}, where BEDT-TTF stands for bis-ethylenedithio-tetrathiafulvalene (in short: ET) and $X = MM^\prime$(SCN)$_4$ a monovalent anion with $M$ = (Rb,Cs,Tl), $M^\prime$ = (Co,Zn). The resulting metastable state, labeled a charge glass (CG), is characterized by frozen short-range charge correlations without long-range order, as shown for instance by NMR \cite{Chiba2008, Sato2017}, x-ray \cite{Kagawa2013,Sasaki2017} and optical conductivity measurements \cite{Hashimoto2014,Sasaki2017}. The observed time-temperature-transformation diagram describing the crystallization and vitrification of charges in these systems suggests that the same nucleation and growth processes that characterize conventional glass-forming liquids guide the crystallization of electrons \cite{Sasaki2017}. 
Furthermore, a phase-change memory function can be realized, enabling the reversible switching between the low-resistive (metastable) CG state and the high-resistive charge-crystal (i.e.\ CO) state with the charge-liquid (CL) state at temperatures above $T_{\text{CO}}$ as a reset state \cite{Oike2015,Kagawa2017}.
The CG-forming ability was found to be a consequence of charge frustration due to the geometric arrangement of the ET molecules on a triangular lattice, which in turn reduces the critical cooling rate $|q_{\rm{c}}|$ required for avoiding the first-order transition \cite{Sato2014a}. The degree of frustration can be quantified by the ratio of the inter-site Coulomb repulsions $V_p/V_c$ along the different crystallographic axes, see Supplemental Material \cite{SI}, which depends on the specific anion \cite{Kagawa2017,Sasaki2017}. Therefore, the most strongly frustrated compounds with $MM^\prime$ = CsCo and CsZn (denoted as \thCsCo\ and \thCsZn)  {\it always}  
exhibit a CG state (lack of CO) on experimental time scales.\\
Recent measurements of resistance fluctuations, a powerful technique to study glassy  
dynamics,  
in \thCsZn\ \cite{Sato2014,Sato2016}, \thRbZn\ \cite{Kagawa2013} and \thTlZn\ \cite{Sasaki2017} have shown that charge clusters exhibit extremely slow and heterogeneous fluctuations when approaching the CG transition temperature from above. Besides the fact that (i) the charge-ordering transition, for instance in \thRbZn, is accompanied by pronounced structural changes \cite{Mori1998,Watanabe2004,Alemany2015} and (ii) the organic charge-transfer salts in general exhibit a strong electron-phonon coupling \cite{Toyota2007}, those results have been interpreted in purely electronic terms.
Theoretical studies, however, have shown that in particular the ET molecules' ethylene endgroups have a strong impact on the CO transition \cite{Alemany2015}, suggesting a peculiar mechanism for the metal-to-insulator transition, namely an order-disorder structural transition of the ethylene endgroups  giving rise to charge localization on non-equivalent ET molecules. Indeed, the question of how these lattice degrees of freedom affect the mechanism of charge crystallization and vitrification remains an important open issue \cite{Sato2014,Sato2017,Sasaki2017}, and therefore it is essential to consider the involvement of structural dynamics in the CG formation.\\ 
In this Letter, we address these questions by presenting thermal expansion measurements on \thCsCo\ and \thCsZn\ combined with cooling-rate-dependent resistance measurements and studies of the charge carrier kinetics by using fluctuation (noise) spectroscopy.  
In particular, the combination of these methods allows to study both the static and dynamic aspects of the glass transition.
We find clear evidence for a structural glass-like transition at $T_{\text{g}}(q) \sim 90-100\,$K which is accompanied by slow dynamics of the charge carriers, well described by a model of non-exponential kinetics. We assign this transition to the freezing of the ET's ethylene groups' conformational motion coupled to the electronic degrees of freedom. This finding challenges the current understanding of the CG formation in the family of $\theta$-(ET)$_2$$X$ compounds.

Single crystals of $\theta$-(ET)$_2$$MM^\prime$(SCN)$_4$ have been grown by electrochemical crystallization \cite{Mori1998}. Electrical contacts were made by using carbon paste and 10 or 25\,$\mu$m-thick gold wires. Resistance measurements of \thCsCo\ were performed along the crystallographic $b$-axis, i.e.\ perpendicular to the conducting ET layers, whereas \thCsZn\ was measured along the in-plane $c$-axis \cite{Note1}. 
Measurements of the resistance fluctuations on \thCsCo\ were performed using a four-terminal DC setup (see \cite{JMueller2011,JMueller2018} for more detailed information). A constant current is applied to the sample and the resistance fluctuations become detectable as voltage fluctuations that are amplified before being processed by a signal analyzer, which calculates the power spectral density (PSD) $S_V(f)$ of the voltage fluctuations. This quantity usually scales with the applied voltage squared, so that the normalized PSD $S_V/V^2 \equiv S_R/R^2$, typically taken at $f = 1\,$Hz, can be used to compare measurements at various temperatures and resistance, $R$, values.
The thermal expansion measurements were carried out by using an ultrahigh-resolution capacitive dilatometer (built after \cite{Pott1983}) enabling the detection of length changes $\Delta L \geq 10^{-2}$ \AA. Thermal expansion and resistance fluctuations in \thCsCo\ have been performed on the same sample, whereas two different samples of \thCsZn\ have been used for thermal expansion and cooling-rate-dependent resistance measurements.

\begin{figure}[t]
	\centering
	\includegraphics[width=1\linewidth]{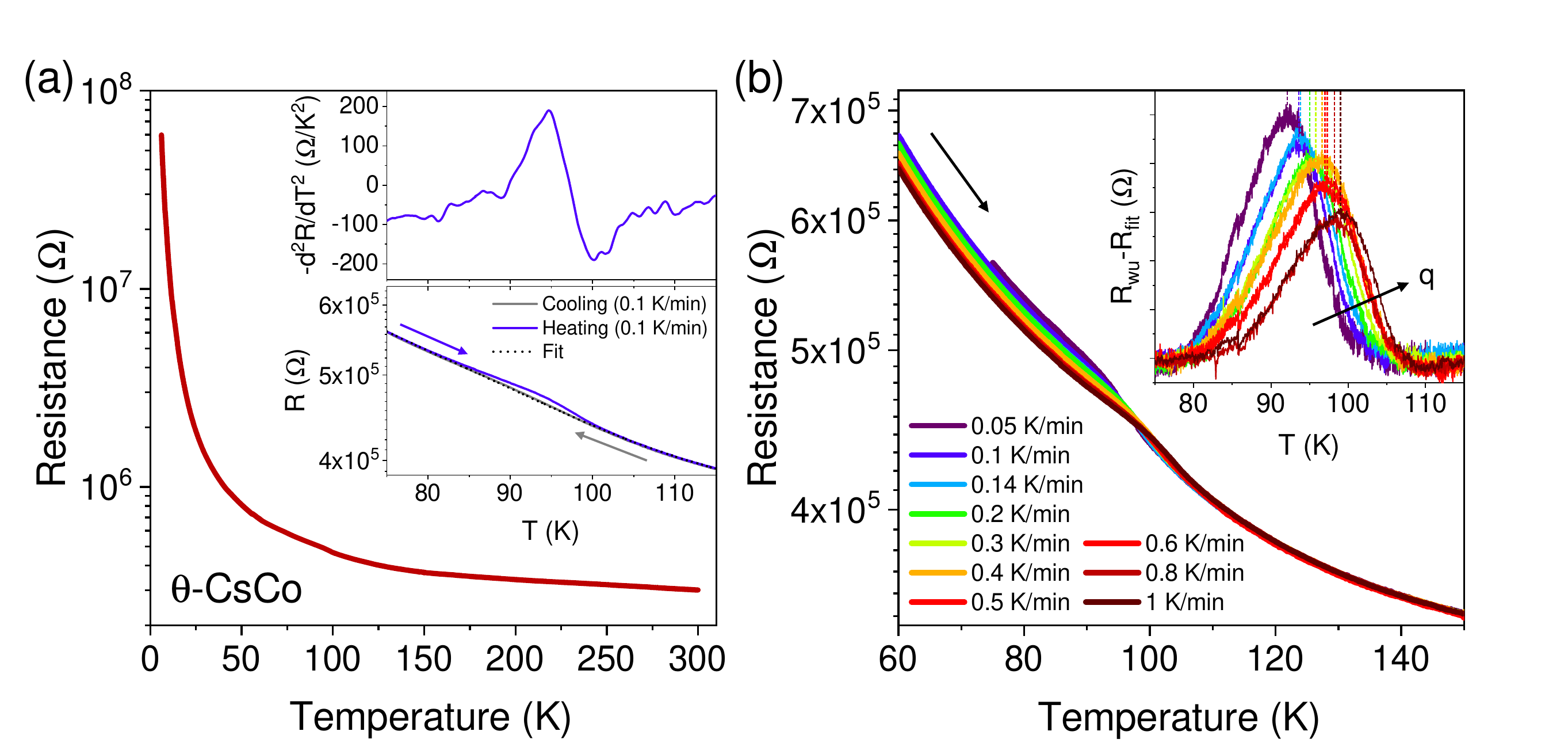}
	\caption{(a) Resistance of \thCsCo\ measured perpendicular to the conducting layers vs.\ temperature for cooling down with $q=-0.8\,$K/min. The insets show the negative second resistance derivative 
	revealing an anomaly at $T_{\rm g} \approx 95\,$K and a hysteresis between warming and cooling. (b) Cooling-rate dependence of the resistance (shown are only warming curves) revealing a shift of the anomaly's temperature (inset).}
	\label{fig:CsCo_res}
\end{figure}

The temperature-dependent resistance of \thCsCo\ for cooling down the sample with $q_{\rm cd} = -0.8\,$K/min is shown in Fig.\ \ref{fig:CsCo_res}(a). Due to the strong frustration of its triangular lattice, this compound always exhibits a continuous change from a CL to a CG state on experimental time scales, i.e.\ for $|q| \gtrsim 0.01$\,K/min, and therefore lacks a CO transition.
At $T_{\text{g}} \sim 95\,$K \cite{Note3} the curve reveals an anomaly, visualized in the second derivative of the resistance (upper inset of Fig.\ \ref{fig:CsCo_res}(a)), which is characterized by a hysteresis between warming and cooling, as depicted in the lower inset of Fig.\ \ref{fig:CsCo_res}(a) for $|q_{\rm cd}| = 0.1\,$K/min. Results of the cooling-rate-dependent resistance are shown in Fig.\ \ref{fig:CsCo_res}(b), which contains only the warming curves. The resistance, which was measured for cooling down with different rates ($q_{\rm cd} = -0.05 \ldots -1\,$K/min) and warming up with $q_{\rm wu} = -q_{\rm cd}$ \cite{SI}, decreases for faster initial cooling, in accordance with previous results on \thCsZn\ \cite{Sato2014}. 
We define the anomaly at $T_{\rm{g}}$ as the maximum in the difference of the resistance curves for warming ($R_{\text{wu}}$) and cooling with $q_{\rm wu} = -q_{\rm cd}$, analyzed by fitting the warming curve excluding the anomaly region with a polynomial function ($R_{\text{fit}}$, cf.\ lower inset in Fig.\ \ref{fig:CsCo_res}(a)) and taking $R_{\text{wu}}-R_{\text{fit}}$. We observe a shift of $T_{\rm{g}}$ to higher temperatures for larger $|q_{\rm cd}|$, as shown in the inset of Fig.\ \ref{fig:CsCo_res}(b). This observation, in combination with the hysteretic behavior, are strong indications of a glassy transition at $T_{\rm{g}}(q) \sim 90 - 100\,$K as has been discussed previously in terms of a frozen charge-cluster glass \cite{Sato2014}. The quantitative analysis of the cooling-rate-dependent resistance is shown below in Fig.\ \ref{fig:expansion}(e) and is discussed together with results of thermal expansion measurements.

The linear coefficient of thermal expansion measured along the $c$-axis, $\alpha_c(T)= {\rm d}\ln{L_c}/{\rm d}T$, is shown in Fig.\ \ref{fig:expansion} for \thCsCo\ (a) and \thCsZn\ (c), revealing a large step-like anomaly at $T_{\rm{g}} \approx 90\,$K \cite{Note3} for both compounds upon slow cooling/warming ($|q_{\rm cd}|=|q_{\rm wu}| = 1.5\,$K/h). We observe a pronounced hysteresis between cooling (shown in blue) and heating (red) with characteristic under- and overshoot behavior in the warming curve typical for a {\it structural} glass-like transition \cite{Gugenberger1992,Nagel2000,JMueller2002,JMueller2004}. At $T_{\rm{g}}^\dagger \approx120\,$K, a smaller feature (marked by arrows) can be recognized for both compounds which is also accompanied by hysteretic behavior.
\begin{figure}[t]
	\centering
    \includegraphics[width=1\linewidth]{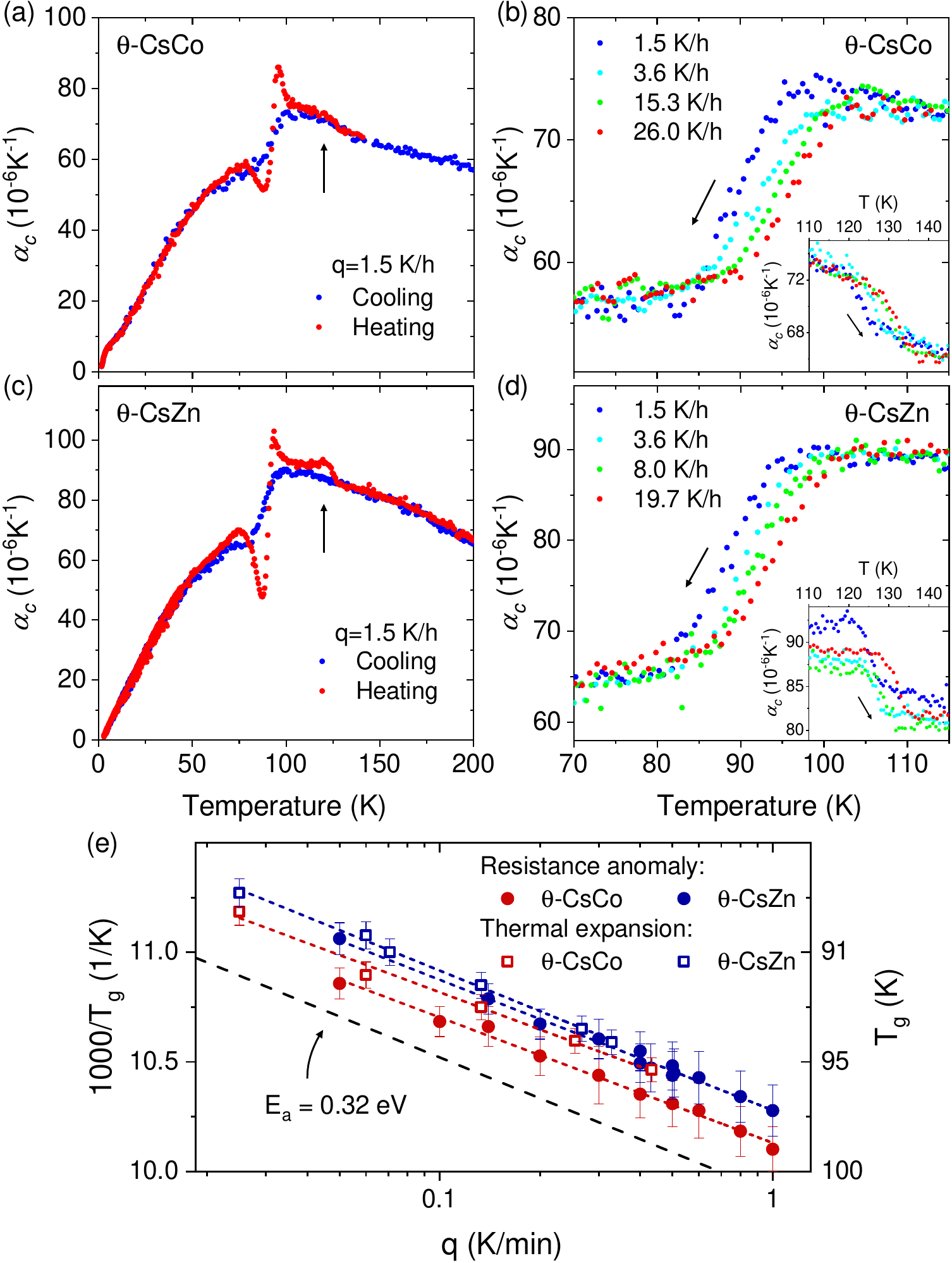}
	\caption{Thermal expansion coefficient $\alpha_c(T)$ of \thCsCo\ and \thCsZn\ measured along the $c$-axis: (a),(c) hysteresis between measurements during warming and cooling; (b),(d) cooling-rate dependence of the anomalies at 90\,K and 120\,K (insets). (e) Arrhenius plot of the cooling-rate-dependent glass transition temperature at $T_{\rm{g}}(q) \sim 90-100\,$K extracted from thermal expansion (squares) and resistance measurements (circles) (see Fig.\ \ref{fig:CsCo_res}(b)), yielding an activation energy of $E_{\rm{a}}=0.32-0.35\,$eV. The black dashed line represents the energy extracted from measurements of resistance fluctuations (see below).}
	\label{fig:expansion}
\end{figure}
The cooling-rate dependence of the thermal expansion coefficient is shown in Fig.\ \ref{fig:expansion}(b) and (d) in the region of the anomalies at $\sim 90-100$\,K and at $\sim 120-130$\,K (inset), revealing a shift of the glass-like transition, usually defined as the midpoint of the step-like feature in the cooling curve, to higher values for larger $|q_{\rm cd}|$. The anomaly at $T_{\rm{g}}^\dagger \approx 120$\,K was analyzed by considering the step in the heating curve due to a rather smooth feature upon cooling. 
The variation of $T_{\rm{g}}^{-1}$ with the cooling rate (on a $\log$-scale) is shown in Fig.\ \ref{fig:expansion}(e) for \thCsCo\ (red squares) and \thCsZn\ (blue squares), indicating a thermally activated relaxation time $\tau=\tau_0\exp[E_{\rm{a}}/(k_{\rm{B}}T)]$ assuming $-|q|\cdot \left.\frac{d \tau}{dT} \right|_{T_{\rm{g}}}\approx 1$ as a defining criterion for the glass-transition temperature \cite{Cooper1982}, with $|q| = |q_{\rm cd,wu}|$. 
In addition, the cooling-rate dependence of the resistance anomaly extracted from the curves shown in Fig.\ \ref{fig:CsCo_res}(b) and from resistance measurements on \thCsZn\ (not shown) is plotted (circles), matching well with the results from thermal expansion measurements. A linear fit to the data sets according to an Arrhenius law, $\ln|q|=-E_{\rm{a}}/(k_{\rm{B}}T_{\rm{g}})+\rm{const.}$ \cite{JMueller2002}, is represented by dashed lines yielding an activation energy $E_{\rm{a}} = 0.32 - 0.35\,$eV for both samples and different measurement techniques. Thus, we conclude that the resistance anomaly and the anomaly at $T_{\rm{g}}(q) \sim 90 -100\,$K seen in thermal expansion measurements are of the same origin. Applying the same analysis to the anomaly at $T_{\rm{g}}^\dagger(q) \sim 120 - 130\,$K yields an activation energy of $E_{\rm{a}}^\dagger=(0.42\pm0.03)\,$eV (not included in Fig.\ \ref{fig:expansion}(e)), which will be discussed later. The step-like thermal expansion anomaly at $T_{\rm{g}} \approx 90\,$K is very similar to and occurs in the same temperature region as the structural glass-like transition in the $\kappa$-(ET)$_2X$ salts \cite{JMueller2002,JMueller2004} where it was assigned to the freezing of the ET's terminal ethylene groups in the energetically unfavored (eclipsed or staggered) configuration. Therefore, we ascribe the anomaly at $T_{\rm{g}}$ in \thCsCo\ and \thCsZn\ to the same origin, which implies that CG formation for these $\theta$-(ET)$_2X$ compounds is intimately linked to or even caused by a glassy structural transition. In Ref.\ \cite{Alemany2015}, it has been argued that in $\theta$-CsZn and $\theta$-CsCo only one of the ethylene endgroups is thermally disordered at high temperatures thereby preventing a metal-to-insulator transition accompanied by CO  upon lowering the temperature. This would imply that only the disordered ethylene moieties show a glass-like freezing, similar to recent observations and calculations for $\kappa$-(ET)$_2$Hg(SCN)$_2$Cl \cite{Gati2018}, where only one of the two crystallographically inequivalent ethylene endgroups freeze in a glassy manner.

\begin{figure}[b]
	\centering
	\includegraphics[width=1\linewidth]{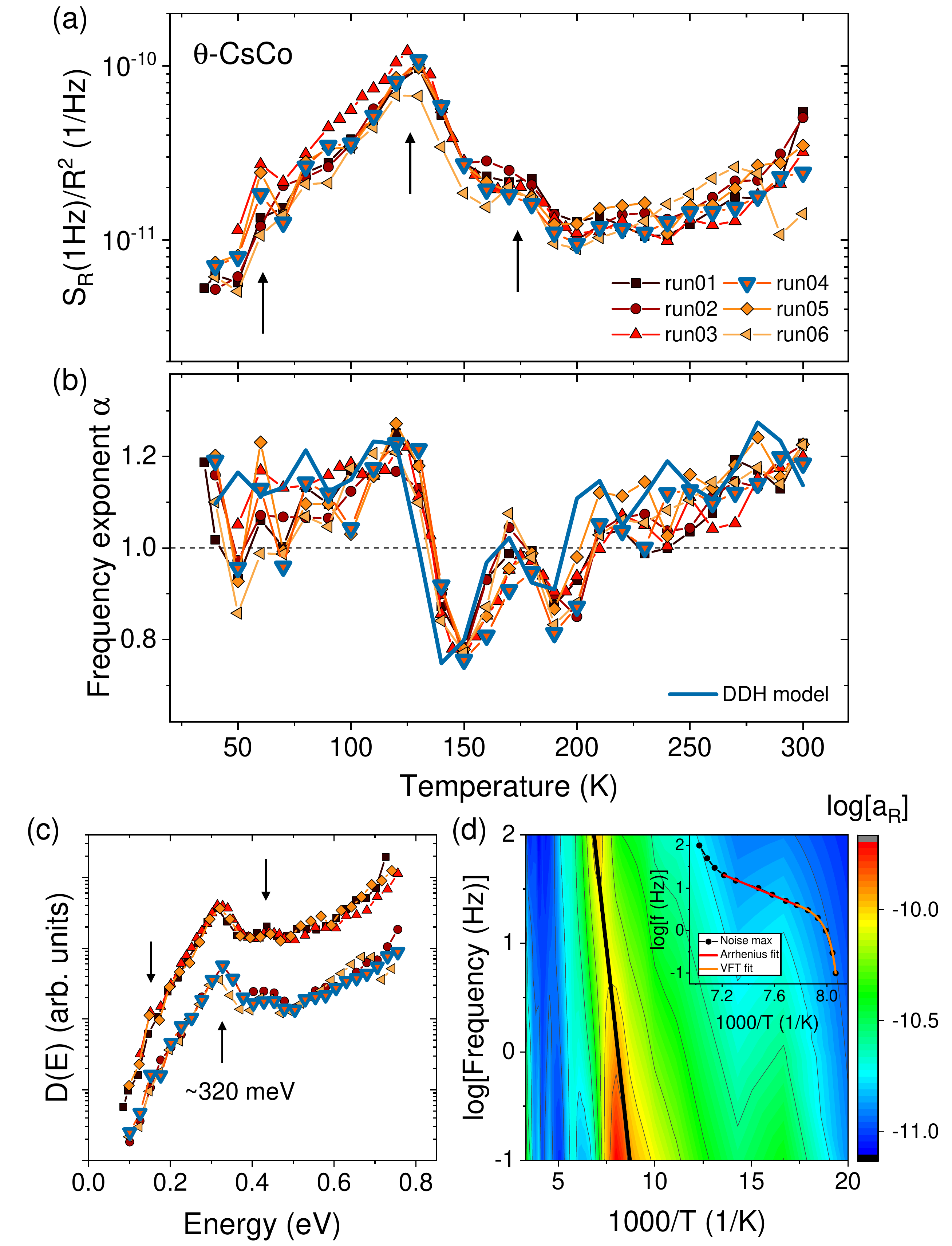}
	\caption{(a) Normalized PSD of the resistance fluctuations $S_R/R^2$ taken at 1\,Hz in \thCsCo, where different colors denote repeated measurement runs and (b) frequency exponent with the calculated $\alpha$ (blue line) after the DDH model, Eq.\ (\ref{alpha_DDH}), for the fourth run. (c) Energy distribution according to the DDH model, Eq.\ (\ref{eq:3}), showing a maximum at 320\,meV. (The shift between the two groups of data sets is due to the use of two different parameters $b = -3.3$ and $-3.7$.) 
	(d) Contour plot of the relative noise level (run03) vs.\ inverse temperature vs.\ logarithmic frequency. The black line represents the energy extracted from the DDH model. The inset shows the frequency-dependent noise maximum, which is fitted by an Arrhenius law (red line) and a Vogel-Fulcher-Tammann law (orange line).}
	\label{fig:CsCo_noise}
\end{figure}

In order to gain insight into the low-frequency dynamics of charge carriers, fluctuation spectroscopy has proven to be a sensitive tool and strong changes are expected when a glassy freezing of electronic or --- due to the electron-lattice coupling --- structural degrees of freedom occurs. Measurements of the resistance fluctuations have been performed in discrete temperature steps during warming or cooling and reveal pure $1/f^\alpha$-noise spectra from room temperature down to the lowest measured temperatures, see \cite{SI}. The normalized resistance noise PSD, $S_R/R^2$, of \thCsCo\ taken at $f =1$\,Hz shown in Fig.\ \ref{fig:CsCo_noise}(a) exhibits a pronounced global maximum at $T\approx130\,$K which is accompanied by a strong increase of the frequency exponent $\alpha$ from 0.8 to 1.2 upon decreasing temperature, as shown in Fig.\ \ref{fig:CsCo_noise}(b), rather similar to the behavior observed in the $\kappa$-(ET)$_2X$ salts due to the glassy freezing of the ET molecules' ethylene endgroup degrees of freedom \cite{JMueller2015}. In addition, at $T\approx175\,$K and $60\,$K small but significant shoulders (marked by arrows) are visible, the former also being accompanied by a corresponding shift of spectral weight to lower frequencies, i.e.\ a slowing down of the dynamics. Repeated measurements (different runs are marked by different colors and symbols) yield very similar results, demonstrating an excellent reproducibility of the observed spectral features in the charge fluctuations.
The temperature-dependent resistance noise can be analyzed in terms of a weighing function of activation energies, related to the distribution of spectral weight of the charge fluctuations, by applying the phenomenological model by Dutta, Dimon and Horn (DDH) \cite{Dutta1979}, which assumes a superposition of --- {\it a priori} not specified ---  independent two-level fluctuators. Their distribution of activation energies $D(E)$ then causes a characteristic temperature dependence of $S_R/R^2(T)$ and a non-monotonic behavior in the frequency exponent $\alpha(T)$ that reflects the shape of $D(E)$ such that for $\alpha$ greater or smaller than 1, $\partial D(E)/\partial E >0$ or $\partial D(E)/\partial E <0$, respectively. In the generalized DDH model it is
\begin{equation}
\frac{S_R(f)}{R^2}(T)=\int_{0}^{\infty}g(T)\frac{4\tau}{1+4\pi^2 f^2\tau^2}D(E)dE,
\end{equation}
where the function $g(T)$ takes into account an explicit temperature dependence of the energy distribution \cite{Black1983,Fleetwood1984,Raquet1999} which can describe a coupling of the fluctuating entities to the measured resistance and --- assuming a power law $g(T) = aT^b$ --- simply causes a vertical offset of the frequency exponent, see e.g.\ \cite{JMueller2012,JMueller2015}. The frequency exponent in the DDH model \cite{Dutta1979,Raquet1999} is then given by
\begin{equation}
\alpha_{\text{DDH}}(T)=1-\frac{1}{\ln(2\pi f\tau_0)}\left[\frac{\partial \ln \frac{S_R(f)}{R^2}(T)}{\partial \ln T} - b -1\right].
\label{alpha_DDH}
\end{equation}
When matching the experimental observations, the model allows to predict the distribution of spectral weight merely from the temperature dependence of the noise magnitude.
The calculated values, indicated by the blue line in Fig.\ \ref{fig:CsCo_noise}(b), are exemplarily shown for the fourth measurement run with $\tau_0=10^{-13.5}\,$s, a typical inverse phonon frequency, and $b = -3.3$. The temperature-dependent $\alpha_{\text{DDH}}$ agrees very well with the experimental curves and even some small features are reproduced. This implies that the model's assumptions are valid, which in turn allows for the calculation of the energy distribution \cite{Dutta1979,Raquet1999} via
\begin{equation}
D(E) \propto  \frac{2\pi f}{k_{\text{B}}T} \frac{S(f,T)}{T^b}.
\label{eq:3}
\end{equation}
\noindent The distribution of activation energies $D(E)$ shown in Fig.\ \ref{fig:CsCo_noise}(c), with the energy of the fluctuating entities $E = k_BT \ln{(2 \pi f \tau_0)^{-1}}$ being derived from the thermal energy and a large logarithmic factor \cite{Kogan1996}, reveals a pronounced peak at $E\approx320\,$meV, which --- for $f = 1$\,Hz --- is associated with the global noise maximum at $130\,$K. The smaller feature at about $430\,$meV is related to  the local maximum of the temperature-dependent PSD at $175\,$K, and the shoulder at $60\,$K corresponds to the small increase at $150\,$meV (marked by arrows). The noise analysis based on the DDH model agrees very well with previous results on $\theta$-CsZn \cite{Sato2016}, as might be expected considering the close vicinity of both systems in the phase diagram of the $\theta$-(ET)$_2$X salts \cite{Mori1998}. 
However, in \cite{Sato2016,Sato2020} the results have been interpreted in purely electronic terms without taking structural dynamics into account. Since the energy of $E\approx320\,$meV extracted from the noise analysis coincides with the energy determined from the cooling-rate-dependent thermal expansion and resistance measurements, we can assign the enhanced noise magnitude at $T\approx130\,$K and the shift of spectral weight to lower frequencies to slow dynamics caused by the glass-like freezing of the ET's ethylene endgroups which undergo a static glassy transition at $T_{\rm{g}}(q) \sim 90-100$\,K \cite{Note2}.
Also, the energy of the noise feature at $175\,$K ($E^\dagger=430\,$meV) matches very well the activation energy of the cooling-rate-dependent anomaly at $T_{\rm{g}}^\dagger(q) \sim 120-130\,$K in the thermal expansion coefficient, which implies that the slowing down of charge dynamics at $175\,$K is related to the second glass-like transition at  $T_{\rm{g}}^\dagger$ seen in thermal expansion measurements. 
Notably, this temperature coincides with (i) a minimum in the in-pane resistivity, see \cite{SI}, indicating a crossover from metallic to semiconducting behavior of the 2D-confined electron fluid \cite{Sato2020} and (ii) the development of a superlattice structure observed by x-ray diffuse scattering \cite{Sato2014} characterized by the wave vector ${\bf {\it q}}_1 \sim (2/3,k,1/3)$, which has been interpreted as growth and subsequent freezing of charge clusters upon cooling.\\
The main features of the resistance fluctuations become highlighted in a contour plot of the dimensionless relative noise level $a_{\rm{R}} = f \times S_R(f,T)/R^2$, see Fig.\ \ref{fig:CsCo_noise}(d), in dependence of inverse temperature and frequency (exemplarily shown for the third run). Notably, a closer look at the frequency dependence of the global noise maximum reveals a more complex behavior than a simple Arrhenius law with a single activation energy of $E\approx320\,$meV (black line). Therefore, the noise magnitude was analyzed for different frequencies ($0.1 - 100$\,Hz), see \cite{SI}. The frequency (on a $\log$-scale) at which the maximum occurs in dependence of the inverse temperature is displayed in the inset of Fig.\ \ref{fig:CsCo_noise}(d). The extracted points show unusually strong deviations from an Arrhenius law upon approaching $T_{\rm{g}}$.
Such a curvature often is described by the Vogel-Fulcher-Tammann (VFT) equation, which is commonly used to determine the viscosity (or time constant) of glass formers above $T_g$. According to a VFT fit, which is represented by the orange line (inset of Fig.\ \ref{fig:CsCo_noise}(d)), this would indicate an extremely fragile glass-forming system where the energy barrier strongly changes with temperature, in contrast to the notion of a strong glass former suggested in \cite{Sato2016}. Also, the noise peak shows an upturn to higher frequencies at higher temperatures, which is highly unusual and not seen for the structural glass-like transition of the terminal ethylene groups in the Mott insulator/superconductor $\kappa$-(ET)$_2$Cu[N(CN)$_2$]$Z$ with $Z$ = Cl,Br \cite{JMueller2015}. 
The data for $\theta$-CsZn reported in Ref.\ \cite{Sato2016} (albeit not discussed explicitly) indicate a similar behavior of unconventional glassy dynamics. This might be a signature of the entanglement of slow structural and charge-cluster dynamics due to the intimate coupling of lattice and electronic degrees of freedom.

In conclusion, the combination of thermal expansion and resistance fluctuation measurements on the molecular conductors \thCsCo\ and \thCsZn\ reveals the static and dynamic characteristics of a structural glass-like transition which we assign to the configurational degrees of freedom of the ET's ethylene endgroups. Our finding discloses an important aspect of the generalized phase diagram of the $\theta$-(ET)$_2X$ family and renders the current understanding of the CG state incomplete:
at least for the present systems which lack a CO transition, the glassy structural dynamics must be taken into account. It naturally raises the question to what extent slow structural dynamics is also involved in the less frustrated systems $\theta$-RbZn or $\theta$-TlZn exhibiting a CO transition that can be kinetically avoided. A future challenge will be to distinguish/separate the glassy characteristics caused by the charges and the molecular entities in order to determine the driving force of the crystallization and vitrification of electrons in a frustrated lattice.\\

We acknowledge support by the Deutsche Forschungsgemeinschaft (DFG, German Research Foundation) through TRR 288 - 422213477 (projects A06 and B02).
This work was also supported by Grants-in-Aid for Scientific Research (KAKENHI) from MEXT, Japan (No. JP21H01793, JP20H05144, JP19H01833, and JP18KK0375), and Grant-in-Aid for Scientific Research for Transformative Research Areas (A) “Condensed Conjugation” (No. JP20H05869) from Japan Society for the Promotion of Science (JSPS). Y.S. and M.L. acknowledge technical assistance by S.\ Hartmann.

T.T. and Y.S. contributed equally to this work.

\bibliographystyle{apsrev4-2}

\end{document}


\title{Involvement of structural dynamics in the charge-glass formation in molecular metals}

\author{Tatjana Thomas}
\affiliation{Institute of Physics, Goethe University Frankfurt, 60438 Frankfurt (M), Germany}
\author{Yohei Saito}
\affiliation{Institute of Physics, Goethe University Frankfurt, 60438 Frankfurt (M), Germany}
\author{Yassine Agarmani}
\affiliation{Institute of Physics, Goethe University Frankfurt, 60438 Frankfurt (M), Germany}
\author{Tim Thyzel}
\affiliation{Institute of Physics, Goethe University Frankfurt, 60438 Frankfurt (M), Germany}
\author{Kenichiro Hashimoto}
\affiliation{Department of Advanced Materials Science, University of Tokyo, 277-8561 Chiba, Japan}
\affiliation{Institute for Materials Research, Tohoku University, 980-8577 Sendai, Japan}
\author{Takahiko Sasaki}
\affiliation{Institute for Materials Research, Tohoku University, 980-8577 Sendai, Japan}
\author{Michael Lang}
\affiliation{Institute of Physics, Goethe University Frankfurt, 60438 Frankfurt (M), Germany}
\author{Jens M\"uller}
\email[Email: ]{j.mueller@physik.uni-frankfurt.de}
\affiliation{Institute of Physics, Goethe University Frankfurt, 60438 Frankfurt (M), Germany}

\date{\today}

\begin{abstract}
\vspace{0.5cm}
\begin{center}
\LARGE{Supplementary Information}
\end{center}
\end{abstract}

\maketitle

\subsection{Crystal structure}

\begin{figure}[h]
	\centering
	\renewcommand{\figurename}{FIG. S$\!\!$}
	\includegraphics[width=1\linewidth]{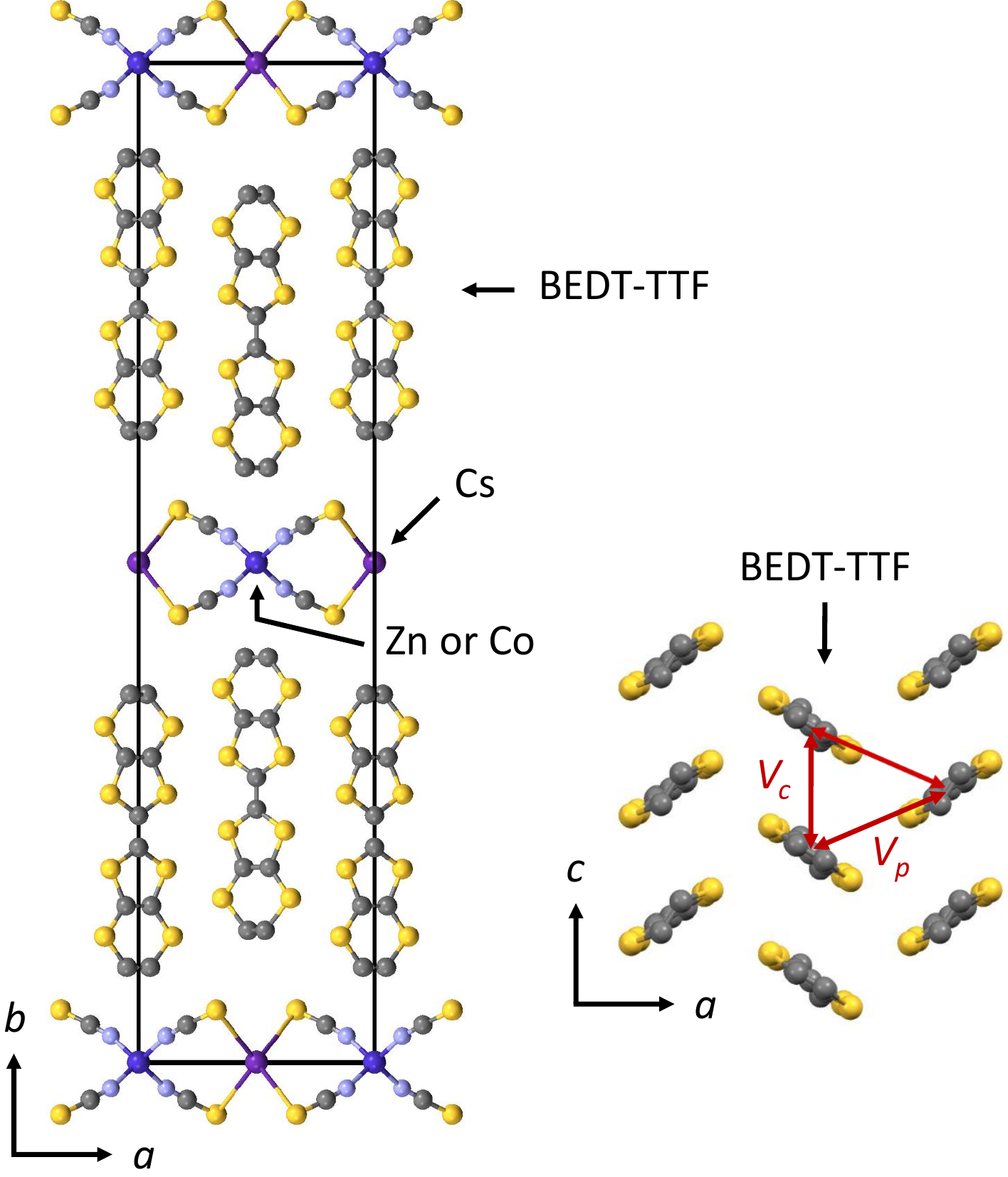}
	\caption{Crystal structure of $\theta$-(BEDT-TTF)$_2$Cs$M^\prime$(SCN)$_4$ with $M^\prime$=(Co,Zn) viewed along different crystallographic axes. The molecular metals consist of conducting and insulating layers (left), where the black rectangle marks the unit cell. The right figure shows the arrangement of the donor molecules on a triangular lattice with the relevant inter-site Coulomb repulsions $V_c$ and $V_p$.}
	\label{fig:SM0}
\end{figure}

The crystal structure of $\theta$-(BEDT-TTF)$_2$Cs$M^\prime$(SCN)$_4$ with $M^\prime$=(Co,Zn) is shown in Fig.\ S\ref{fig:SM0}. The BEDT-TTF molecules form two-dimensional conducting layers, which are separated by insulating sheets consisting of the anions Cs$M^\prime$(SCN)$_4$ with $M^\prime$=(Co,Zn) (Fig.\ S\ref{fig:SM0} (left)). The arrangement of the donor molecules on a triangular lattice (Fig.\ S\ref{fig:SM0} (right)) leads to geometric frustration, when the ratio of the inter-site Coulomb repulsion $V_p/V_c$ is close to unity.

\subsection{Determination of glass-transition temperature and activation energy from cooling-rate-dependent resistance measurements}
\begin{figure}[h]
	\centering
	\renewcommand{\figurename}{FIG. S$\!\!$}
	\includegraphics[width=1\linewidth]{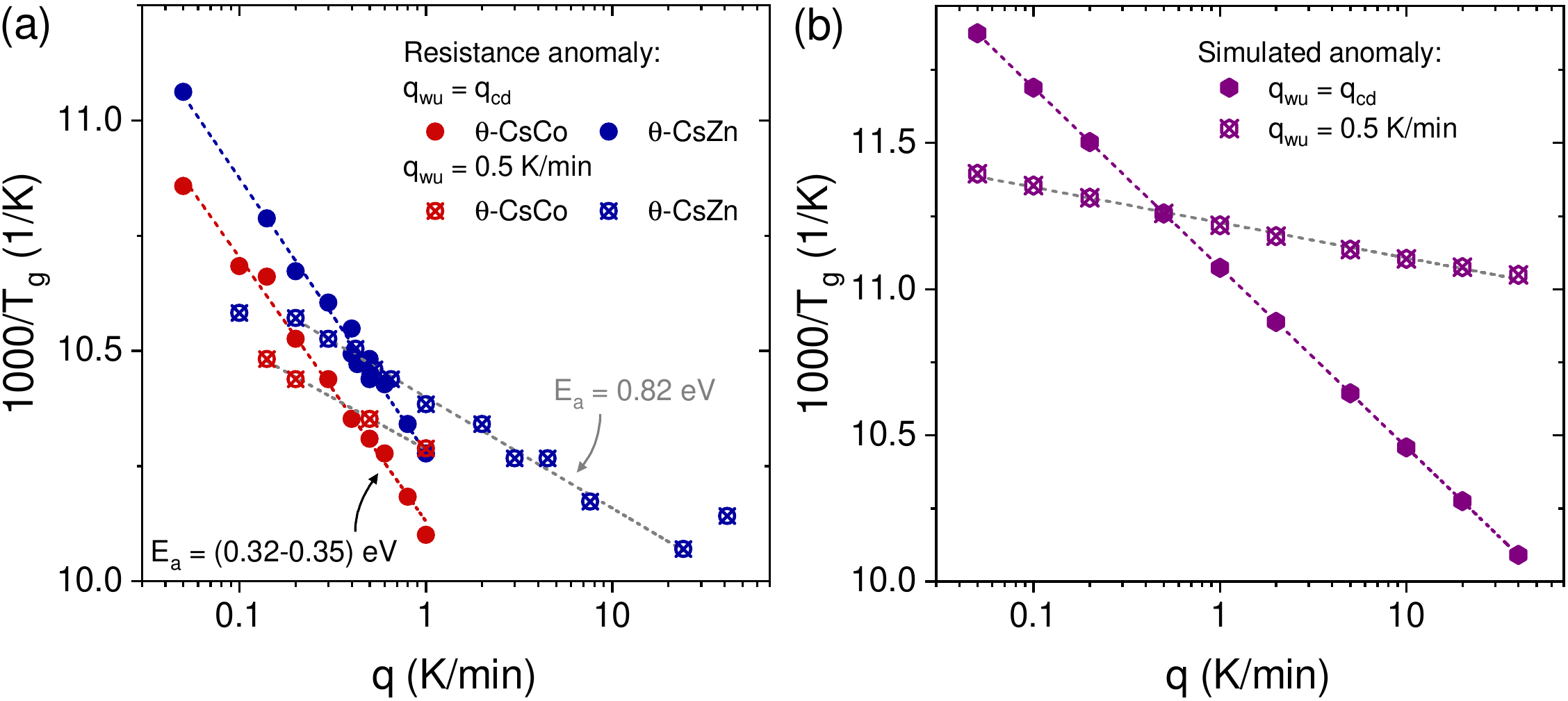}
	\caption{Arrhenius plot of the glass-transition temperature at $T_{\rm{g}}\approx90-100\,$K in dependence of cooling rate: (a) Experimental results of the resistance anomaly for $q_{\text{wu}}=-q_{\text{cd}}$ (filled circles) and for a fixed warming rate $q_{\text{wu}}=0.5\,$K/min after cooling down with different rates $q_{\text{cd}}=- 0.05 \ldots - 40$\,K/min (circles with crosses). (b) Numerical simulations of the glass-transition temperature for the different measurement methods analyzed from the probability distribution of a two-level system, see \cite{Hartmann2014}, for different cooling rates.}
	\label{fig:SM1}
\end{figure}

Studies of the cooling-rate-dependent resistance have been performed by cooling down and warming up the sample with the same rates of temperature change $|q|$ with $q = {\rm d}T/{\rm d}t$. We find that another procedure which is commonly used \cite{Wang2002,Fischer2018}, namely cooling down the sample at a range of different rates (e.g.\ $q_{\rm cd} = -0.05$\,K/min til $-40$\,K/min and warming up with a {\it fixed} rate (e.g.\ $q_{\text{wu}}^{\text{fixed}} = 0.5\,$K/min) may lead to misleading results.\\ 
The results of both experimetal methods are shown in Fig.\ S\ref{fig:SM1}(a), where the inverse glass-transition temperature is plotted in dependence of the cooling rate (on a $\log$-scale). The comparison reveals strong differences in $T_{\text{g}}$ and thus in the determined activation energy (given by the slope of the dashed and dotted lines). At the specific cooling rate $q_{\text{cd}} = - q_{\text{wu}}^{\text{fixed}}$, the curves for both measurement protocols intersect. We interpret the results as follows: Upon slow heating after fast cooling, $T_{\text{g}}$ is underestimated whereas for faster warming than previous cooling, $T_{\text{g}}$ is overestimated, resulting in a flatter slope of the curve. Therefore, when using a constant warming rate after cooling with different rates, there might be a significant error due to the early or delayed relaxation of the quenched metastable configurations when approaching the glass-transition temperature.

For numerical simulations of the glass-transition temperature we used a relaxation model of glassy dynamics as described in \cite{Hartmann2014} in order to determine the probability distribution of a simple two-level process for different cooling rates. By calculating the occupation probability $p_1(T)$ of the lower energy level in the double-well potential for cooling and heating with given rates, where we have used the parameters $E_{\text{a}}=0.32\,$eV and $\Delta E=200\,$K for the double-well potential and $\nu_0=10^{15}\,$Hz for the attempt frequency, we define the glass-like anomaly as the maximum in the difference of both curves, $p_{\text{1,wu}}-p_{\text{1,cd}}$. The analysis for the two different protocols is shown in Fig.\ S\ref{fig:SM1}(b). Again, we observe strong differences in the determined glass-transition temperature with qualitative behavior similar to our experimental findings. These results demonstrate the importance of the correct choice of thermal protocol for the analysis of the glassy transition in the present systems.

\subsection{Resistance noise spectroscopy}
\begin{figure}[h]
	\centering
	\renewcommand{\figurename}{FIG. S$\!\!$}
	\includegraphics[width=1\linewidth]{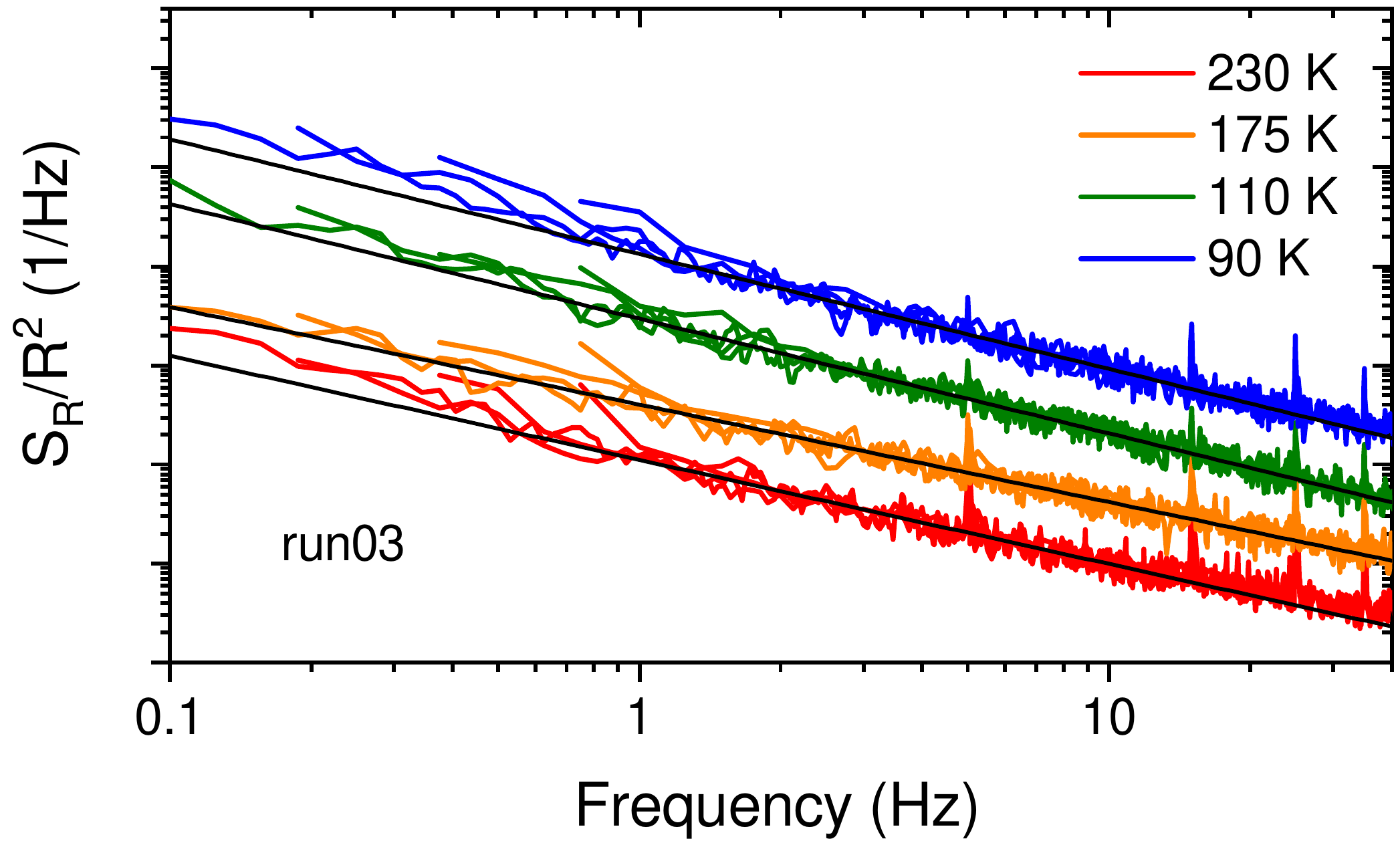}
	\caption{Typical spectra of resistance fluctuations for selected temperatures in a double-logarithmic plot. The curves are shifted for clarity; black lines correspond to linear fits.}
	\label{fig:SM3}
\end{figure}
The normalized power spectral density of the resistance fluctuations in dependence of frequency, $S_R/R^2(f)$, is shown in Fig.\ S\ref{fig:SM3} for selected temperatures, where the curves are shifted for clarity. The spectra reveal pure $1/f^{\alpha}$-behavior (i.e.\ without superimposed Lorentzian contributions \cite{JMueller2020}) from room temperature down to the lowest measured temperature. From linear fits in this $\log$-$\log$ representation, we determine the noise magnitude $S_R/R^2(T,f = 1\,{\rm Hz})$ and the frequency exponent $\alpha(T) \equiv - \partial \ln{[S_R/R^2(f,T)]}/\partial \ln{f}$ discussed in the main paper.\\

\begin{figure}[h]
	\centering
	\renewcommand{\figurename}{FIG. S$\!\!$}
	\includegraphics[width=1\linewidth]{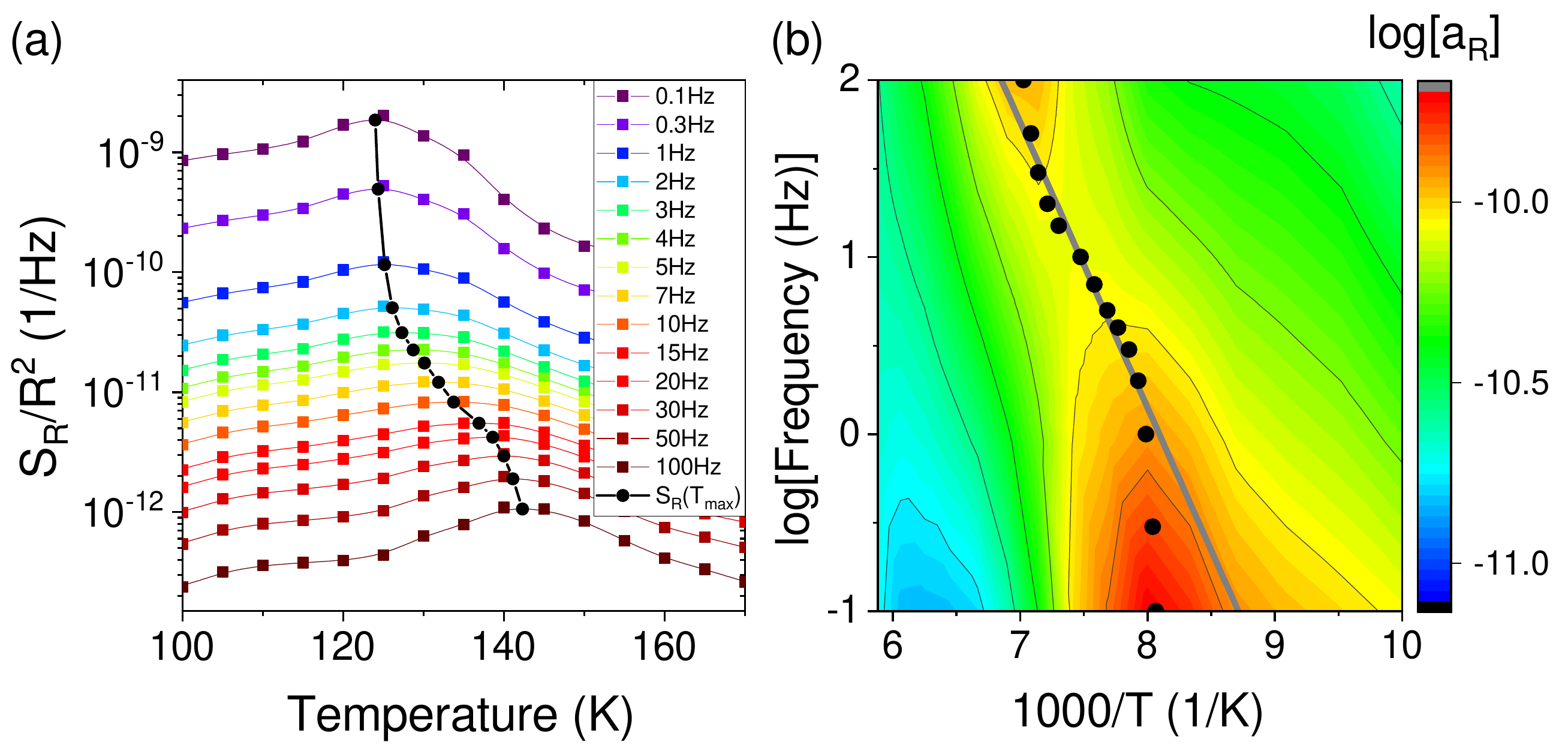}
	\caption{(a) Noise magnitude analyzed for different frequencies in dependence of temperature. The black circles mark the frequency-dependent noise maximum around $T\sim130\,$K. (b) Zoom of the contour plot of the relative noise level including the frequency-dependent noise maximum (black circles) and the energy extracted from the DDH model (gray line) described in the main paper.}
	\label{fig:SM2}
\end{figure}
The noise magnitude in dependence of temperature, $S_R/R^2(T,f={\rm const.})$, was analyzed for different frequencies ($0.1-100\,$Hz), as shown in Fig.\ S\ref{fig:SM2}(a), where the maximum is indicated by black circles (lines are merely guides for the eyes). This frequency-dependent noise maximum is added to the contour plot of the relative noise level (Fig.\ S\ref{fig:SM2}(b)) in dependence of frequency (on a $\log$-scale) and inverse temperature (cf.\ Fig.\ 3(d) in the main paper), revealing unusually strong deviations from a simple Arrhenius behavior (gray line).

\subsection{In-plane resistivity}
\begin{figure}[h]
	\centering
	\renewcommand{\figurename}{FIG. S$\!\!$}
	\includegraphics[width=1\linewidth]{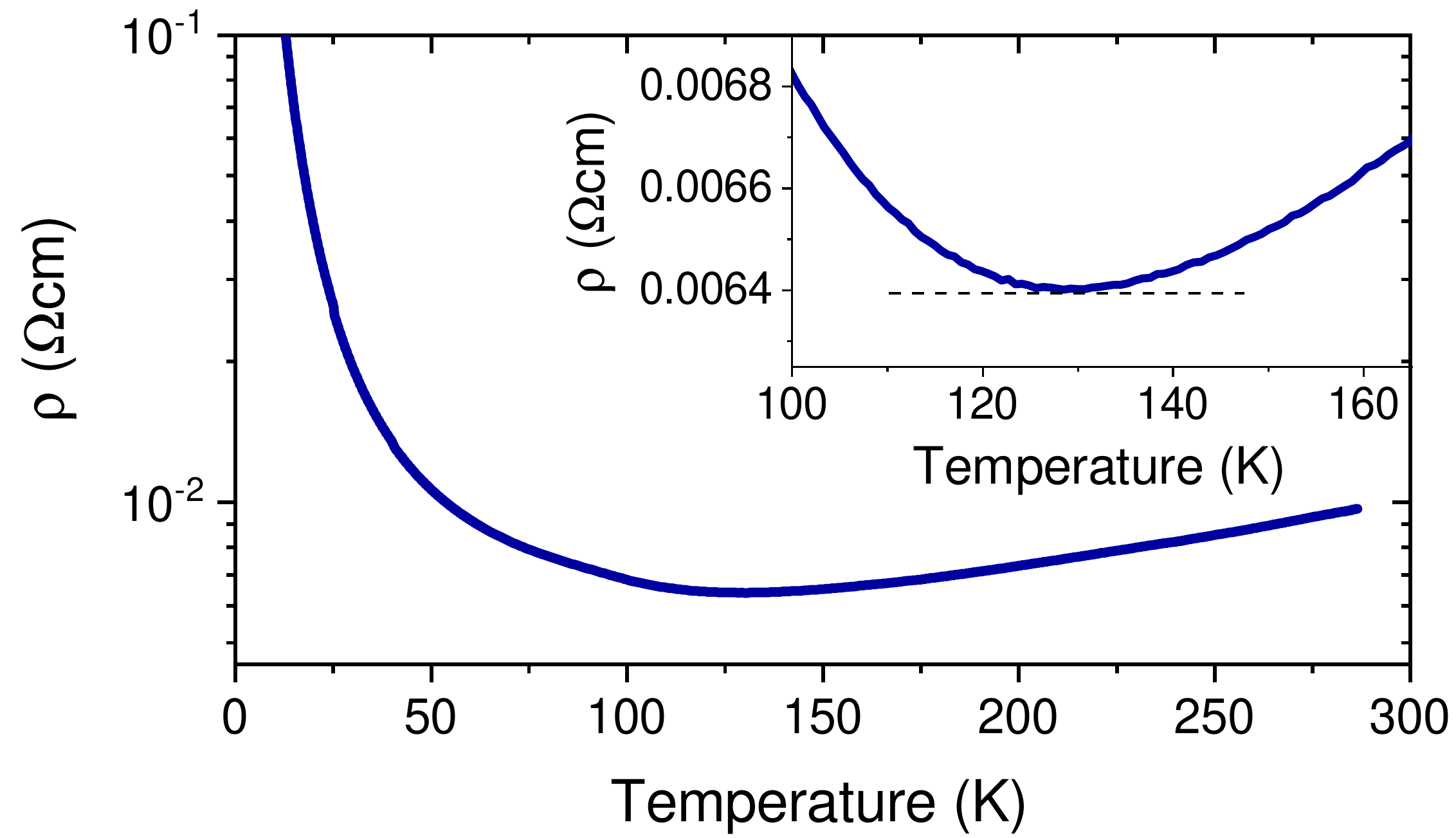}
	\caption{In-plane resistivity of $\theta$-CsZn for a cooling rate $q = - 0.5$\,K/min. Inset emphasizes the temperature range around $\sim 120 - 130$\,K where the minimum in the resistivity coincides with the glass-like transition at $T_g^\dagger$ observed in the thermal expansion, see main paper.}
	\label{fig:SM5}
\end{figure}
The in-plane resistivity of $\theta$-CsZn is shown in Fig.\ S\ref{fig:SM5} for a cooling rate $q = -0.5$\,K/min. The inset highlights a pronounced minimum which for different samples is consistently observed at $T \sim 120 - 130$\,K , thereby coinciding with the second glass-like transition at $T_{\rm{g}}^\dagger$ observed in thermal expansion. For higher cooling rates, the minimum seems to both shift and become broader (not shown), which makes the thermal expansion results shown in the main paper a more suitable to determine the activation energy.
